\documentclass[twocolumn, traditabstract]{aa} 
\usepackage{graphicx}
\usepackage{amsmath} 
\usepackage{subeqnarray}
\usepackage{layouts}
\usepackage{color}
    \definecolor{Blue}{rgb}{0.0,0.0,1.0}
    \definecolor{Red}{rgb}{1.0,0.0,0.0}
    \definecolor{Green}{rgb}{0.0,0.0,1.0}

\definecolor{battleship}{rgb}{0.77,0.75,0.81}

\newcommand{\be}{\begin{equation}}
\newcommand{\ee}{\end{equation}}
\newcommand{\bea}{\begin{eqnarray}}
\newcommand{\eea}{\end{eqnarray}}

\begin{document}
\title{Coronal ejection and heating in variable-luminosity X-ray sources}
\author{W{\l}odek Klu\'zniak}
\institute{Nicolaus Copernicus Astronomical Center, ul. Bartycka 18,
  PL-00-716 Warszawa, Poland \\ \email{wlodek@camk.edu.pl}
}
\date{Received August 1, 2012; accepted ????}
\abstract{A sudden increase in stellar luminosity may lead to the ejection
of a large part of any optically thin gas orbiting the star.
Test particles in circular orbits will become unbound, and will escape
to infinity (if radiation drag is neglected), when  the
luminosity changes from zero to at least one half the Eddington value,
or more generally, from $L$ to $(L_{\rm Edd}+L)/2$ or more.
Conversely, a decrease
in luminosity will lead to the tightening of orbits of optically thin
fluid. Even a modest fluctuation of luminosity of accreting
neutron stars or black holes is expected to lead to substantial
coronal heating.
Luminosity fluctuations may thus account for the high temperatures of
the X-ray corona in accreting black hole and neutron star systems.}

\keywords{Accretion disks -- Scattering -- X-rays: binaries --
 Stars: winds, outflows -- Stars: neutron}
\maketitle

\section{Introduction}

Low mass X-ray binaries (LMXBs) are bright X-ray sources whose
luminosity is thought to be powered by accretion from a binary companion
onto a neutron star or a black hole. Typically, each source can be found
in one of two or more distinct spectral states, some of which include a hard
X-ray power law component interpeted as radiation from a hot
($\sim 10^2\,$keV) corona. 
Each state has an associated characteristic variability on many timescales.
Nearly all sources exhibit quasi-random excursions of luminosity.
In this \emph{Letter} I examine some of the consequences of rapid
variations of luminosity, particularly for optically thin Keplerian
flow, which I identify with the X-ray corona.

Since the paper of \cite{Walker} it is understood that high luminosity
of an accreting neutron star will be associated with removal of
angular momentum from the optically thin flow via the Poynting--Robertson
drag. In fact, it seems to be a part of the lore of neutron star
astrophysics that the primary effect of high luminosity is to bring
matter down from its nearly-circular orbits towards the inner parts of
the accretion disk, and perhaps even onto the stellar surface.
This is ideed true for optically thin flow in a steady radiation field.
However, a rapid change in luminosity may have the opposite
effect and, as I show below, under certain conditions
may lead to ejection of the optically thin matter.

An even more important effect will occur when the luminosity
undergoes a small but rapid change. The optically thin fluid will
then continue its orbital motion in very nearly the same circular orbits,
but will undergo strong heating, caused by dissipation of excess kinetic
energy. This may explain the presence of hot X-ray coronae in some 
spectral states of LMXBs.
\section{Keplerian orbits}
In Newtonian physics, computing the response of  an orbiting test particle
to a sudden change in luminosity is a simple exercise in orbital mechanics.
Consider orbits in a spherically symmetric potential
\be{V(r)=-\frac{GM}{ r}}.
\label{pot}
\ee
As shown by Newton these are conic sections with eccentricity
\be
e=\sqrt{1+\frac{2El^2}{G^2M^2}},
\label{ecc}
\ee
where $E$ and $l$ are the specific (per unit mass) energy and angular
momentum of the orbiting particle.
The specific kinetic energy and angular momentum
 of a particle in circular orbit
of radius $r_0$ are
      \be{K_0=\frac{GM}{2 r_0}},\ \ l_0=\sqrt{GMr_0},
       \label{circorbit}
      \ee
and the total specific energy is
      \be{E_0=-\frac{GM}{2 r_0}.}
      \ee
Suppose that the energy of the particle is changed to $E$, with no change
in  its angular momentum. The particle will now move in an orbit with 
eccentricity
      \be{e=\sqrt{1+E/K_0}.}
      \ee
In particular, $E=E_0 \Rightarrow e=0$, i.e., the original orbital energy
corresponds to a circular orbit; while $E=0  \Rightarrow e=~1$,
i.e., a marginally unbound particle
moves in a parabolic orbit. It is well known that
 $E>0$ corresponds to an unbound hyperbolic orbit.
As we will see, the effect of an impulsive change of luminosity on
test particles in circular orbits is equivalent to a change of the particle
energy.

\section{Luminosity effects}
Optically thin hydrogen plasma suffers radiation pressure forces
proportional to the Thomson cross-section and the radiative flux.
At Eddington luminosity, $L_{\rm Edd}$, the radiative force balances
gravity exactly at any radius---in Newtonian physics of a spherically
symmetric source both  the radiative flux
$L/(4\pi r^2)$, and the force of gravity $-GM/r^2$ drop off inversely with
the square of the radial distance. Hence, at lower luminosity the presence
of radiation pressure is equivalent
to a proportional reduction of the gravitational mass by irradiation:
$GM$ may simply be replaced by $GM(1-\lambda)$,
with $\lambda\equiv L/L_{\rm Edd}$, so that eqs.~(\ref{pot}), (\ref{ecc})
become
\be
  V(r)=-\frac{GM(1-\lambda)}{r},
\ee
\be
  e=\sqrt{1+\frac{2El^2}{G^2M^2(1-\lambda)^2}}.
\label{ecclum}
\ee
The specific kinetic energy
in circular orbit at $r_0$ for a source with luminosity
$L$ and true gravitational mass $M$ is
\be K_\lambda={\frac{GM(1-\lambda)}{2 r_0}}.
\ee

Upon an impulsive change of stellar luminosity from zero to $L$,
a particle that has been travelling travelling
in circular orbit at $r_0$ 
would conserve its kinetic energy and angular momentum,
$K_0$,  $l_0$ of
Eq.~(\ref{circorbit}), so that its new
orbit would be described by the following specific energy and angular momentum:

      \be{E=\frac{GM}{2 r_0}-\frac{GM(1-\lambda)}{r_0},
        \ \ \ l=l_0=\sqrt{GMr_0}},
      \ee
and because of the excess energy, $K_0-K_\lambda=\lambda GM/(2r_0)$,
this orbit would no longer be circular.
The eccentricity of the orbit is, from Eq.~(\ref{ecclum}),
      \be\large{e=\frac{\lambda}{1-\lambda}}.
      \ee
In particular, $\lambda\ge 1/2 \Rightarrow e\ge 1$, i.e., if the luminosity
is now one half of the Eddington value, or greater, the particle is in an
unbound orbit. If, on the other hand, the increase was from zero
to less than one half of the Eddington luminosity, the particle remains
in  a bound elliptic orbit, $\lambda< 1/2\Rightarrow e<1$,
 with an increased semi-major axis $a=r_0(1-\lambda)/(1-~2\lambda)$.

\section{Coronal ejection}
\label{ejection}

In reality, the fluctuations of luminosity in LMXBs do not occur between
$L=0$ and $L\ne0$, but rather between some initial $L_1\ne0$ and some
final value $L\ne0$. Generalizing the derivation from the previous section,
we start with circular orbits at luminosity $L_1$, with
$\lambda_1\equiv L_1/L_{\rm Edd}$, the specific kinetic energy and
angular momentum in
circular orbits in the optically thin region being given by
\be
 K_1=\frac{GM(1-\lambda_1)}{2 r}, \ \ \ l_1=\sqrt{GM(1-\lambda_1)r}.
\ee
Following an impulsive change of luminosity to $L$, the new orbits
are defined by a new potential and the same values of kinetic energy
and angular momentum:
\be
V(r)=-\frac{GM(1-\lambda)}{r},
\ \ \ K=K_1, \ \ \ l=l_1,
\ee
where, as before,
$ \lambda\equiv L/L_{\rm Edd}$.
The eccentricity of the new orbits is
\be
   e=\frac{|\lambda-\lambda_1|}{1-\lambda}.
\ee

Identifying the corona with the optically thin region, the
condition for its ejection ($ e\ge~1$) now becomes:
      \be{1-\lambda\le (1-\lambda_1)/2}
      \ee
i.e.,
\be L_{\rm Edd} - L \le (L_{\rm Edd}-L_1)/2,
\ee
or
\be
    (L_{\rm Edd}+L_1)/2\le L.
\label{ejection}
\ee
Thus, the condition for coronal ejection is that the luminosity reduces
its distance to the Eddington value by at least a factor of two.
Clearly, the closer the initial value $L_1$ is to $L_{\rm Edd}$,
the lower the fractional value of luminosity increase necessary
for the ejection of the corona. E.g., an increase from $0.8L_{\rm Edd}$
to  $0.9L_{\rm Edd}$ corresponds to a fluctuation of less than $13\%$,
and this is sufficient to clear out the corona.

\section{Coronal heating}
\label{coronal}
Now we turn to changes of luminosity that do not lead to coronal ejection.
This could be because the luminosity has increased (from $L_1$) by less than
$(L_{\rm Edd}-L_1)/2$, thus violating condition~(\ref{ejection}), or because
the luminosity has decreased. In any case, the new test particle orbit
is an ellipse with semi-major axis $r_0(1-\lambda)/|1+\lambda_1-2\lambda|$,
in the notation of the previous Sections. The periastron is at radius
$r_0$ for $\lambda>\lambda_1$, and at
\be
  r_-=r_0(1-\lambda_1)/(1+\lambda_1-2\lambda),
\label{peri}
\ee
for $\lambda<\lambda_1$. For a small decrease of luminosity, the latter
value becomes $r_-\approx r_0[1-2(\lambda_1-\lambda)/(1-\lambda_1)]$.

Naturally, the state of the corona cannot be described by test particle
orbits, as those of different particles would intersect. It is clear that
some energy will be dissipated, and the fluid orbits will circularize.
The ultimate outcome is difficult to predict, because redistribution
of angular momentum may occur. However, it seems clear that the dissipated
energy will be on the order of $(\lambda_1-\lambda)^2$ times the virial energy
of the corona.

Consider the evolution of a geometrically thin annulus,
initially orbiting at $r_0$, under a small impulsive
 change of luminosity of the central star, 
$|\delta\lambda|\equiv |\lambda-\lambda_1|<<1$,
and subsequent dissipation of energy.
If the annulus conserves its angular momentum, it will settle down
in a circular orbit at $r_1=r_0(1-\lambda_1)/(1-\lambda)$
after dissipating an amount
of energy equal to
\be
\frac{(\delta\lambda)^2}{1-\lambda_1}\frac{GM}{2r_0}.
\ee

\section{Discussion}
The observed accreting neutron stars and black holes are usually quite luminous
and are typically variable in time. It seems important to explore the
consequences of rapid changes of luminosity, whose magnitude
 may be a large fraction of the Eddington value.

The relevance of Poynting-Robertson drag to neutron stars was
first pointed out by \cite{Walker}.
While there is no doubt that high luminosity is accompanied by 
radiation drag, which at least in
the optically thin regime in a steady source will eventually remove
kinetic energy and angular momentum  of orbiting matter,
 this process takes time.
If the luminosity undergoes a rapid change, the first and immediate
response of orbiting matter is to change its trajectory.

By considering Newtonian orbital mechanics, I have shown that an
impulsive increase of central luminosity of sufficiently
high magnitude, $L-L_1 > (L_{\rm Edd}-L_1)/2$,
may lead to an ejection of the optically thin corona
on a dynamical timescale.
This may have an application to X-ray spectral-state 
changes of black holes and neutron stars, and may be of some
importance in X-ray bursts. Inclusion of radiation drag requires
numerical computations, and is postponed till another paper,
where a fully general relativistic discussion of the
problem will be presented \cite{Stahl}.

The coronal response to small changes in luminosity was considered
in Section~\ref{coronal}. It seems inevitable that relatively
minor changes in the luminosity of the central source lead to
substantial energy dissipation. This could be the as yet unexplained
mechanism of coronal heating. The estimated magnitude of the effect is
rather large. The dissipated energy at 50 Schwarzschild radii
caused by a single excursion in luminosity of 10\% Eddington
($0.1L_{\rm Edd}$) is on the order
of $10^{-4}$ of the coronal rest mass, i.e., it corresponds to a temperature of
$\sim100\,$keV. At 5 Schwarzschild radii the same result will
be obtained by a fluctuation of only 3\% Eddington luminosity.

In passing, we note that for a steady source of luminosity $\lambda$
in Eddington units, the orbital frequency in the optically thin
regime is modified by a factor of $\sqrt{1-\lambda}$. Care must
be taken to account for this effect
when interpreting the redshift/blueshift of any spectral features
in terms of orbital motion.

\bigskip\par\noindent
The results of Sections 2--4 were presented on October 28, 2011
at the LOFT Science Meeting in Amsterdam.\ \ \
http://www.isdc.unige.ch/loft/index.php/meetings/loft-science-meeting

\end{document}